\documentclass[aps,prl,twocolumn,reprint,superscriptaddress,citeautoscript]{revtex4-2}

\usepackage{physics}
\usepackage{graphicx}
\usepackage{amsmath,amssymb,bm}
\usepackage[normalem]{ulem}

\usepackage{color}

\usepackage{hyperref}
\hypersetup{
	citecolor = blue,
	colorlinks = true,
	urlcolor = blue
}

\begin{document}
\title{Gyromagnetic Angular Momentum Interconversion in Neutron Stars}

\author{Hiroshi Funaki}
\affiliation{Kavli Institute for Theoretical Sciences, University of Chinese Academy of Sciences, Beijing, 100190, China.}

\author{Yuta Sekino}
\affiliation{Interdisciplinary Theoretical and Mathematical Sciences Program (iTHEMS), RIKEN, Wako, Saitama 351-0198, Japan}
\affiliation{Nonequilibrium Quantum Statistical Mechanics RIKEN Hakubi Research Team, RIKEN Cluster for Pioneering Research (CPR), Wako, Saitama 351-0198, Japan}
\affiliation{RIKEN Cluster for Pioneering Research (CPR), Astrophysical Big Bang Laboratory (ABBL), Wako, Saitama, 351-0198 Japan}

\author{Hiroyuki Tajima}
\affiliation{Department of Physics, School of Science, The University of Tokyo, Tokyo 113-0033, Japan}

\author{Shota Kisaka}
\affiliation{Physics Program, Graduate School of Advanced Science and Engineering, Hiroshima University, Hiroshima 739-8526, Japan}

\author{Nobutoshi Yasutake}
\affiliation{Department of Physics, Chiba Institute of Technology (CIT), 2-1-1 Shibazono, Narashino, Chiba 275-0023, Japan}
\affiliation{Advanced Science Research Center, Japan Atomic Energy Agency, Tokai, 319-1195, Japan}

\author{Mamoru Matsuo}
\email{mamoru@ucas.ac.cn}
\affiliation{Kavli Institute for Theoretical Sciences, University of Chinese Academy of Sciences, Beijing, 100190, China.}
\affiliation{CAS Center for Excellence in Topological Quantum Computation, University of Chinese Academy of Sciences, Beijing 100190, China}
\affiliation{Advanced Science Research Center, Japan Atomic Energy Agency, Tokai, 319-1195, Japan}
\affiliation{RIKEN Center for Emergent Matter Science (CEMS), Wako, Saitama 351-0198, Japan}

\begin{abstract}
We propose a novel mechanism for angular momentum (AM) exchange between the crust and core of a neutron star (NS) via the gyromagnetic effect. Using extended hydrodynamics, we model the star by incorporating macroscopic AM and microscopic AM originating from neutron orbital and spin AM.
We reveal that macroscopic dynamics in the crust can inform microscopic AM in the core leading to neutron spin polarization, and offer alternative scenario of (anti-)glitches. 
This work highlights the overlooked 
multi-scale AM interconversions in NS physics, paving the way for gyromagnetic astrophysics. 
\end{abstract}

\maketitle

{\it Introduction.---}
Angular momentum is crucial in neutron star (NS) physics, influencing their formation, evolution, and observable phenomena. During massive star collapse, its conservation leads to rapid rotation, affecting the resulting structure and stability. In NSs, angular momentum plays pivotal role in glitches, caused by unpinning superfluid vortices, and anti-glitches~\cite{Antonopoulou2022}. Recent high-precision observations resolving angular-velocity changes during glitches demand a multi-scale approach, integrating both macroscopic dynamics and microscopic contributions from neutron spin and orbital angular momentum.
However, there remains uncertainty regarding the angular momentum transport process itself inside NSs.

The gyromagnetic effect, a quintessential example of non-equilibrium multi-scale angular momentum dynamics, was discovered by Barnett~\cite{barnett1915Magnetization}, Einstein, and de Haas~\cite{a.einstein1915} in the early 20th century. It involves the interconversion between macroscopic and microscopic angular momentum, originally observed as the exchange between mechanical angular momentum carried by a rotating magnetic body and the spin and orbital angular momentum of electrons. 
Recently, this effect has been found in a wide range of materials~\cite{barnett1935Gyromagnetic,scott1962review}, including magnetic~\cite{Wallis-APL-2006-09,imai2018observation,imai2019angular,Dornes2019} and non-magnetic substances~\cite{ono2015Barnett,ogata2017Gyroscopic,hirohata2018magneto}, elastic bodies~\cite{zolfagharkhani2008nanomechanical,kobayashi2017Spin,kurimune2020Highly,kurimune2020Observation,tateno2020Electrical,Harii-2019-NatCommun,Mori2020,tateno2021Einstein}, liquid metals~\cite{takahashi2016Spin,takahashi2020Giant,tabaeikazerooni2020Electron,tabaeikazerooni2021Electrical,tokoro2022spin}, nuclear spin systems~\cite{chudo2014Observation,chudo2015Rotational,harii2015Line,chudo2021Barnett,wood2017magnetic}, and quark-gluon plasma~\cite{adamczyk2017global,adam2018global,adam2019polarization,acharya2020evidence,adam2021global}, manifesting across a broad spectrum of angular velocities from a few Hz to $10^{21}$ Hz.

\begin{figure}[ht]
 \includegraphics[width=0.95\hsize]{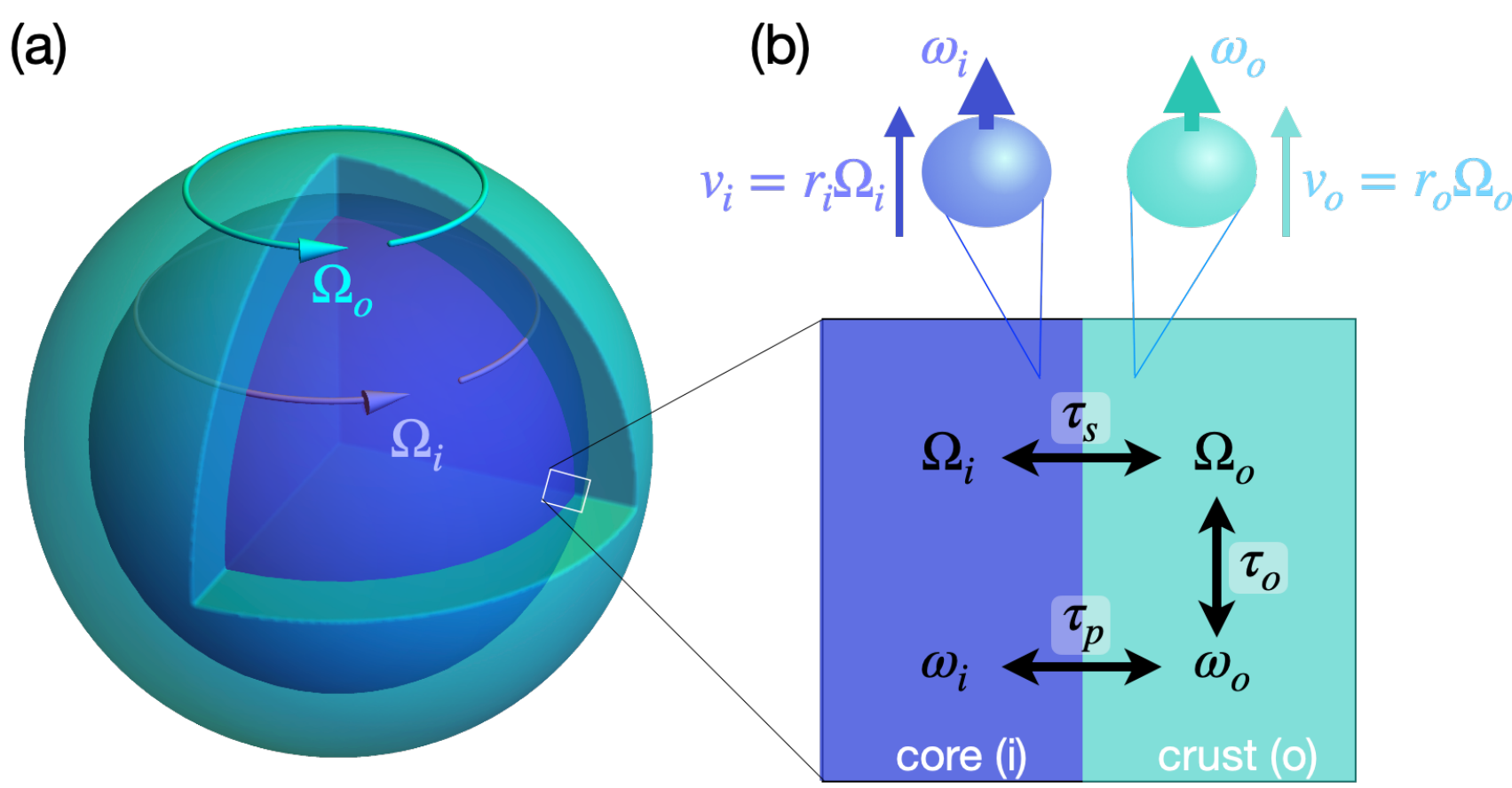}
 \caption{Schematic representation of a gyromagnetic angular momentum interconversion model in a NS. (a) The star is divided into the crust (outer side, o) and core (inner side, i), where macroscopic angular momentum dynamics are represented by angular velocities $\Omega_{\rm o}$ and $\Omega_{\rm i}$. (b) Microscopic angular velocities $\omega_{\rm o}$ and $\omega_{\rm i}$, originating from the spin and orbital angular momenta of constituent neutrons, are also considered. These four angular momenta are coupled through angular momentum relaxation processes ($\tau_{\rm s}$, $\tau_{\rm o}$, and $\tau_{\rm p}$) driven by gyromagnetic effects.}
 \label{fig1}
\end{figure}

The original gyromagnetic effect involved the interconversion between macroscopic mechanical angular momentum and microscopic angular momentum in a rigidly rotating magnetic body, driven by angular-momentum conservation. Recent advancements in spintronics and quark-gluon systems have expanded this concept to include the exchange between macroscopic angular momentum, such as vorticity in elastic bodies~\cite{Wallis-APL-2006-09,zolfagharkhani2008nanomechanical,kobayashi2017Spin,kurimune2020Highly,kurimune2020Observation,tateno2020Electrical,Harii-2019-NatCommun,Mori2020,tateno2021Einstein} or fluids~\cite{takahashi2016Spin,takahashi2020Giant,tabaeikazerooni2020Electron,tabaeikazerooni2021Electrical,tokoro2022spin}, and microscopic angular momentum from electron or quark spins~\cite{adamczyk2017global,adam2018global,adam2019polarization,acharya2020evidence,adam2021global}.
Modern descriptions of the gyromagnetic effect extend continuum mechanics to include \textit{microrotation}~\cite{eringen2012microcontinuum,lukaszewicz2012micropolar}, an intrinsic angular velocity of continuum elements independent of translational motion, departing from conventional fluid dynamics such as the Navier-Stokes equation, which neglects internal rotational effects. In this extended framework, coupled equations describe the dynamics of macroscopic vorticity and microrotation, with rotational viscosity~\cite{de2013non} introduced to facilitate angular-momentum exchange. This model explains angular-momentum transfer from fluid vorticity to electron spin in liquid metals~\cite{matsuo2013Mechanical,matsuo2017Theory,takahashi2016Spin} or spin polarization of quarks and gluons in quark-gluon plasma~\cite{hattori2019fate}.
Independently, rotational viscosity effects have also been investigated theoretically in NSs from the perspective of irreversible thermodynamics~\cite{SandovalVillalbazoGarcíaPercianteColín+2001+269+277}.

In this Letter, we propose an effective model to describe the previously overlooked aspects of microscopic angular momentum in the hydrodynamics of NSs, using the extended hydrodynamics known as the micropolar theory--a robust framework for describing gyromagnetic effects and multi-scale angular-momentum conversion. 
As a first step, we derive a four-angular-velocity model from the micropolar theory by dividing the NS into crust and core, where macroscopic and microscopic angular momenta in both regions can interconvert. 
This model extends the conventional two-angular-velocity framework~\cite{Baym1969-le,Shapiro}, which it recovers in the limit where rotational viscosity vanishes.
As shown in Fig.~\ref{fig1}, the four-angular-velocity model introduces additional relaxation channels: between the vorticity and microrotation in the crust, and between the microrotations of the core and crust.
We then numerically demonstrate three possible gyromagnetic phenomena in the crusts of NSs: (i) generation of microrotation by Barnett effect during post-glitch relaxation, (ii) glitches triggered by the Einstein-de Haas effect, and (iii) post-glitch relaxation by the Einstein-de Haas effect [(iii) is shown only in the supplement material~\cite{supplemental}].
These findings provide deeper insights into spin polarization effects in NSs, offer a novel scenario for gyromagnetic-induced glitches and anti-glitches, and suggest a pathway for accessing core microscopic angular momentum information through crust dynamics. We also discuss the observational implications of these results.

{\it Model.---}
To develop an effective model of gyromagnetic angular momentum interconversion
in the coupled crust-core system as shown in Fig. 1, we start with  coupled extended hydrodynamic equations for the crust ($a={\rm o}$) and the core ($a={\rm i}$)
\begin{align}
\rho_a \frac{\partial {\bm v}_a}{\partial t} &= 
 -\rho_a ({\bm v}_a \cdot \nabla) {\bm v}_a
 -\nabla p_a\nonumber\\
 &+\mu_a \triangle {\bm v}_a
 +\mu_{r,a} \nabla \times ( 2{\bm \omega}_a -\nabla \times {\bm v}_a),\label{hydro1}\\
\rho_a I_a \frac{\partial {\bm \omega}_a}{\partial t}&=
 -\rho_a I_a ({\bm v}_a \cdot \nabla) {\bm \omega}_a\nonumber\\
 &+\eta_a \triangle {\bm \omega}_a
 +2 \mu_{r,a} (\nabla \times {\bm v}_a -2{\bm \omega}_a),\label{hydro2}
\end{align}
where $\rho_a$ is the mass density of the fluid element, ${\bm v}_a$ is the fluid velocity, $p_a$ is the pressure, and ${\bm \omega}_a$ is the microrotation, namely, the internal angular velocity  of the fluid element. The parameters $\mu_a$, $\mu_{r,a}$, $\eta_a$, and $I_a$ represent the shear viscosity coefficient, rotational viscosity coefficient, spin diffusion coefficient, and specific moment of inertia of the fluid element, respectively. We note that $\mu_{r,o} = \mu_r$ and $\mu_{r,i} = 0$ are taken for simplicity, assuming that the core ($a={\rm i}$) is in the deep superfluid phase without viscosities~\cite{PhysRevLett.115.020401}, whereas the crust involves viscous flows due to the interaction effects such as nucleon-nucleon scattering~\cite{flowers1976transport,flowers1979transport}.
Equation~\eqref{hydro1} describes 
the translational motion of fluid elements in the crust ($a={\rm o}$) and core ($a={\rm i}$), respectively. 
The first, second, and third terms on the right-hand side are identical to those in the Navier-Stokes equation, representing the advection term, pressure term, and momentum viscosity term, respectively. 
The last term on the right-hand side represents the rotational viscosity effect,
which arises because fluid elements possess an independent degree of freedom in the form of microrotation $\omega_a$, unlike in conventional fluid dynamics.
This rotational motion of the fluid element, characterized by microrotation, interacts with the vorticity $\nabla \times {\bm v}_a$ generated by the relative motion of surrounding fluid elements, attempting to resolve any discrepancies between them.
In turn, Eq.~\eqref{hydro2} describes the rotational motion, or angular momentum dynamics, of fluid elements in the crust and core. The first term on the right-hand side represents the advection of microrotation, while the second term accounts for the diffusion of microrotation, which microscopically originates from the diffusion of neutron spin and orbital angular momentum. The third term, similar to the final term in Eq. (\ref{hydro1}), represents the rotational viscosity effect.

For simplicity, we assume a spherically symmetric system and hence introduce spherical coordinates $(r, \theta, \phi)$ with the origin at the center shared by the crust and core. 
We then derive the angular velocity equation for this one-dimensional system along the $r$-axis. The crust and core are joined at $r = R_i$, with the crust extending to $r = R_o$. In this spherical coordinate system, the fluid velocity is given by ${\bm v}_a = r_a \sin \theta \, \Omega_a \, {\bm e}_{\phi}$ with a unit vector ${\bm e}_{\phi}$ along the axis of the azimuthal angle $\phi$. Furthermore, we assume uniform angular velocities $\Omega_a$ in both the crust and core regions, with microrotation $\omega_a$ also uniform in each region, while allowing for discontinuities near the interface between them.

By spatially integrating out Eqs.~\eqref{hydro1} and \eqref{hydro2}~\cite{supplemental},
we obtain the following four-angular-velocity model:
\begin{align}
 \frac{\partial \Omega_{\rm o}}{\partial t} &=
  -\frac{1}{\tau_{\rm s}} (\Omega_{\rm o} -\Omega_{\rm i})
  -\frac{1}{\tau_{\rm o}} (\Omega_{\rm o}-\omega_{\rm o}),
\label{Eq_Lome_o}
\\
 \alpha_{\rm s} \frac{\partial \Omega_{\rm i}}{\partial t} &=
  +\frac{1}{\tau_{\rm s}} (\Omega_{\rm o} -\Omega_{\rm i}),
\label{Eq_Lome_i}
\\
 \alpha_{\rm o} \frac{\partial \omega_{\rm o}}{\partial t} &=
  -\alpha_{\rm o} \frac{1}{\tau_{\rm p}} (\omega_{\rm o} -\omega_{\rm i})
  +\frac{1}{\tau_{\rm o}} (\Omega_{\rm o}-\omega_{\rm o}),
\label{Eq_Some_o}
\\
 \alpha_{\rm p} \alpha_{\rm o} \frac{\partial \omega_{\rm i}}{\partial t} &=
  +\alpha_{\rm o} \frac{1}{\tau_{\rm p}} (\omega_{\rm o} -\omega_{\rm i}),
\label{Eq_Some_i}
\end{align}
where 
$\tau_{\rm s}^{-1} = {\tilde \mu}_{\rm oi} \frac{ \frac{3}{2} R_{\rm i}^2 {S_{\rm oi}}  }
         { \rho_{\rm o} V_{\rm o} I_{S, {\rm o}} }$
,
$\tau_{\rm p}^{-1} = {\tilde \eta}_{\rm oi} \frac{{S_{\rm oi}}}{\rho_{\rm o} V_{\rm o} I_{\rm o}}$
,
$\tau_{\rm o}^{-1} = \mu_{r} \frac{ 4 }
         { \rho_{\rm o} I_{S, {\rm o}} }$, 
$\alpha_{\rm s} =
  \frac{\rho_{\rm i} V_{\rm i} I_{S,_{\rm i}}}
       {\rho_{\rm o} V_{\rm o} I_{S,{\rm o}}}$, 
$\alpha_{\rm p} =
  \frac{\rho_{\rm i} V_{\rm i} I_{{\rm i}}}
       {\rho_{\rm o} V_{\rm o} I_{{\rm o}}}$, 
$\alpha_{\rm o} =
  \frac{I_{{\rm o}}}{I_{S,{\rm o}}}$, with 
  $I_{S,{\rm o}} =
  \frac{2}{5} \frac{R_{\rm o}^5 -R_{\rm i}^5}{R_{\rm o}^3 -R_{\rm i}^3}$ and 
$I_{S,{\rm i}} = \frac{2}{5} R_{\rm i}^2$. 
Here, ${\tilde \mu}_{\rm oi}$ and ${\tilde \eta}_{\rm oi}$ are the interface transfer coefficients for microrotation and vorticity, respectively.
These equations focus solely on the post-glitch behavior and do not account for angular momentum dissipation due to dipole magnetic radiation. To include this effect, it is sufficient to simply introduce the corresponding dissipation term into equation (3)~\cite{Baym1969-le,Shapiro}.

{\it Barnett effect in NSs.---}
Let us discuss the {\it Barnett effect} in the crust, i.e., neutron spin and orbital polarization induced by the angular momentum transfer from macroscopic to microscopic scales as a representative example demonstrating that the derived four-angular-velocity model captures dynamics beyond the conventional two-angular-velocity model.
In particular, we demonstrate spin and orbital polarization in the crust, driven by the transport coefficient $\tau_{\rm o}$, which represents angular momentum relaxation linking $\Omega_{\rm o}$ and $\omega_{\rm o}$ as introduced in Eqs.~\eqref{Eq_Lome_o} and \eqref{Eq_Some_o}~\cite{supplemental}.
For a typical value of angular velocity
$\bar{\Omega}_{\rm o}=\,1{\rm Hz}$~\cite{Antonopoulou2022}, the macroscopic angular momentum is 
$\bar{L}_{\rm flow}= \rho_{\rm o} V_{\rm o} I_{S,{\rm o}} \bar{\Omega}_{\rm o} \approx 
 \,2.96 \times 10^{36}$~${\rm kg \cdot m^2 /s}$.
On the other hand, 
the total angular momentum for completely spin-polarized neutrons is 
$\bar{L}_{\rm spin}= \rho_{\rm o} V_{\rm o} \frac{\hbar}{2m} \approx \,8.77 \times 10^{20}{\rm kg \cdot m^2 /s}$.
This large scale separation between $\bar{L}_{\rm flow}$ and $\bar{L}_{\rm spin}$ suggests that only a tiny variation of macroscopic $\bar{L}_{\rm flow}$ significantly changes microscopic spin and orbital angular momentum.

{\it Gyromagnetic angular momentum transfer in NSs.---}
Let us consider the dynamical processes of angular momentum transfer via the gyromagnetic effect with the four-angular-velocity model. The model is
 numerically solved for the following two cases; (i) 
generation of microrotation during the conventional glitch
and (ii) gyromagnetic glitch with large fluid element inertia, both of which are induced by rapid relaxation between vorticity and microrotation.
These two cases have been chosen to be consistent with the previous results, which describe the conventional glitch mechanism as a sudden angular momentum transfer from the core to the crust via the unpinning and re-pinning of vortices~\cite{Shapiro}.
In the conventional two-velocity model, the vortex pinning/unpinning process results in the sudden transfer of angular momentum from the core to the crust. In our framework, the recipient of the core’s angular momentum can be either the global crust matter (\(\Delta \Omega_{\rm o}>0\)) in case (i), aligning with the conventional scenario, or the micro-rotation (\(\Delta \omega_{\rm o}>0\)) in case (ii). In both cases, the resulting behavior of \(\Omega_{\rm o}\) successfully accounts for observed glitches.

In this Letter, we assume NSs with a canonical mass, $M=1.4~M_\odot$, 
and adopt the equation of state by S. Typel et al.~\cite{DD2}. These choices correspond to setting the parameters $\alpha_{\rm s}$ and $\alpha_{\rm p}$ to $\alpha_{\rm s}=50$ and $\alpha_{\rm p}=100$, respectively.
Consequently, during the subsequent relaxation process, the angular velocity decreases in the crust while increasing in the core (see $\Omega_{\rm i}$ and $\Omega_{\rm o}$ in both cases).
The relaxation time parameter between angular velocities $\tau_{\rm s}=50~{\rm days}$ is chosen to be roughly consistent with astronomical observations of glitches, while the relaxation time parameter between microrotations is assumed to have the same value $\tau_{\rm p}=50~{\rm days}$.
The relaxation time parameter between angular velocity and microrotation in crust $\tau_{\rm o}=0.2~{\rm days}$ is chosen to be roughly consistent with  observations of glitch rise time in the latter case (ii)~\cite{supplemental}.
%(see the supplemental material for more details).

%%%%%%%%%%%%%%%%%%%%%%%%%%%%%%%%%%%%%%%%%%%%%%%%%%%%%%%
\begin{figure}[t]
\includegraphics[width=1.0\hsize]{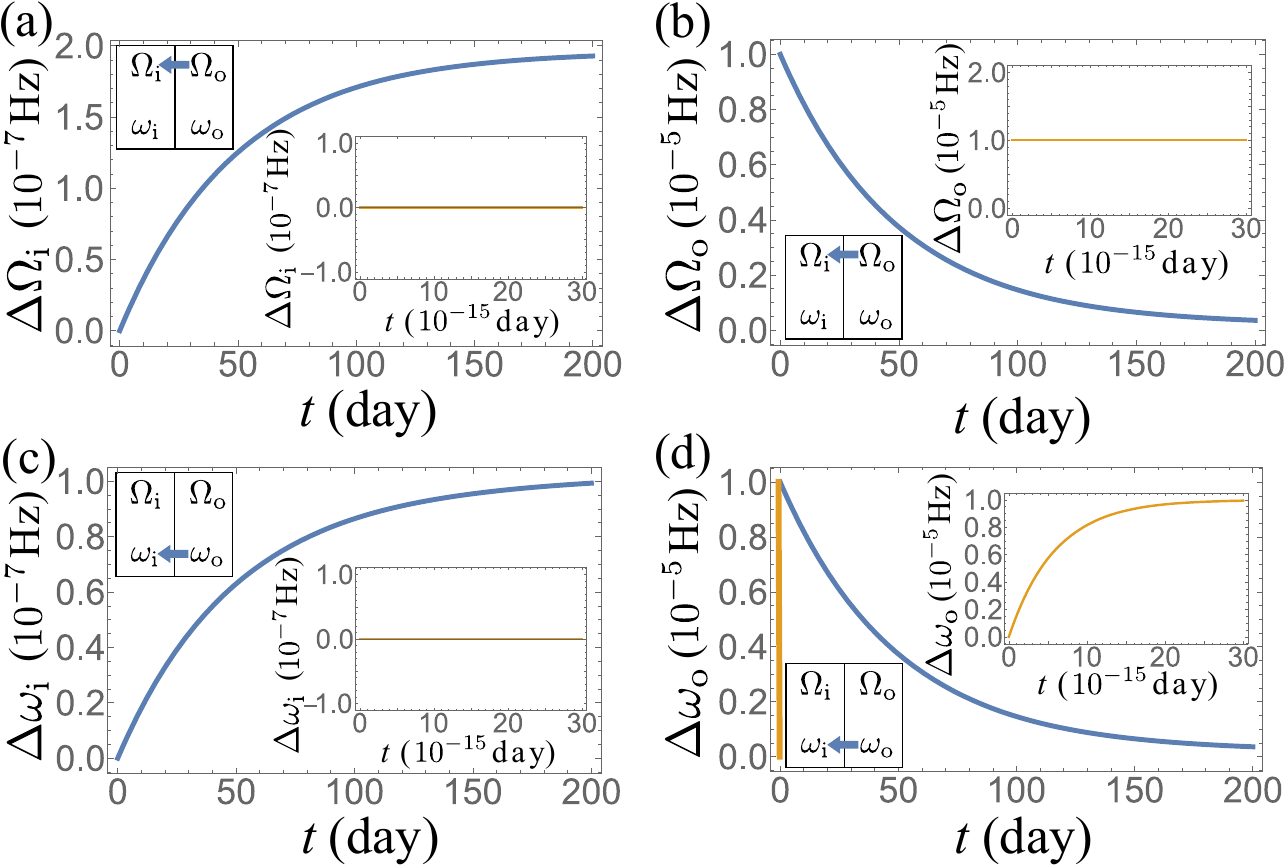}
 \caption{
Generation of microrotation in post-glitch relaxation process with $\alpha_{\rm o}=2.96 \times 10^{-16}$.
(a) Core angular velocity $\Delta\Omega_{\rm i}\equiv\Omega_{\rm i}-\bar{\Omega}_{\rm o}$, (b) crust angular velocity $\Delta\Omega_{\rm o}\equiv\Omega_{\rm o}-\bar{\Omega}_{\rm o}$, (c) core microrotation $\Delta\omega_{\rm i}\equiv\omega_{\rm i}-\bar{\Omega}_{\rm o}$, and (d) crust micrototation $\Delta\omega_{\rm o}\equiv\omega_{\rm o}-\bar{\Omega}_{\rm o}$ from the the reference angular velocity $\bar{\Omega}_{\rm o}=\,1{\rm Hz}$.
The rapid increase of crust microrotation [orange line in (d)] consisting of neutron spin and orbital polarization is triggered by the relaxation of the crust rotation from the initial value $\Delta\Omega_{\rm o}=10^{-5}\,{\rm Hz}$ in (b).
The inset schematically illustrates the flow of angular momentum.
}
 \label{s-IP_G}
\end{figure}
%%%%%%%%%%%%%%%%%%%%%%%%%%%%%%%%%%%%%%%%%%%%%%%%%%%%%%%
%
Fig.~\ref{s-IP_G} shows the fluid dynamics in the case (i), which can be regarded as a straightforward extension of the conventional glitch model: after the unpinning and re-pinning process, the curst suddenly gains angular momentum. Thus, the initial conditions are set as $\Omega_{\rm o} > \Omega_{\rm i} = \omega_{\rm o} = \omega_{\rm i}=\bar{\Omega}_{\rm o}=\,1{\rm Hz}$.
The moment of inertia of the fluid element is set to a small value $\alpha_{\rm o}=2.96 \times 10^{-16}$ to focus on the situation where microrotation does not provide feedback to the angular velocity of the fluid but is expected to affect the polarization of neutron spins~\cite{supplemental}.
The results show the saturated microrotation generation caused by angular momentum transfer from crust vorticity to crust microrotation consisting of neutron spin and orbital angular momentum. %(Fig.~\ref{s-IP_G}).
This process can be regarded as the propagation of microrotations accompanying the transfer of angular momentum, and is thus referred to as the {\it Barnett effect}. %, or simply as the ``post-glitch relaxation process".
This behavior also suggests that there is a significant possibility of complete polarization of neutron spins in the crust driven by the gyromagnetic effect. 

%%%%%%%%%%%%%%%%%%%%%%%%%%%%%%%%%%%%%%%%%%%%%%%%%%%%%%%
\begin{figure}[t]
 \includegraphics[width=1.0\hsize]{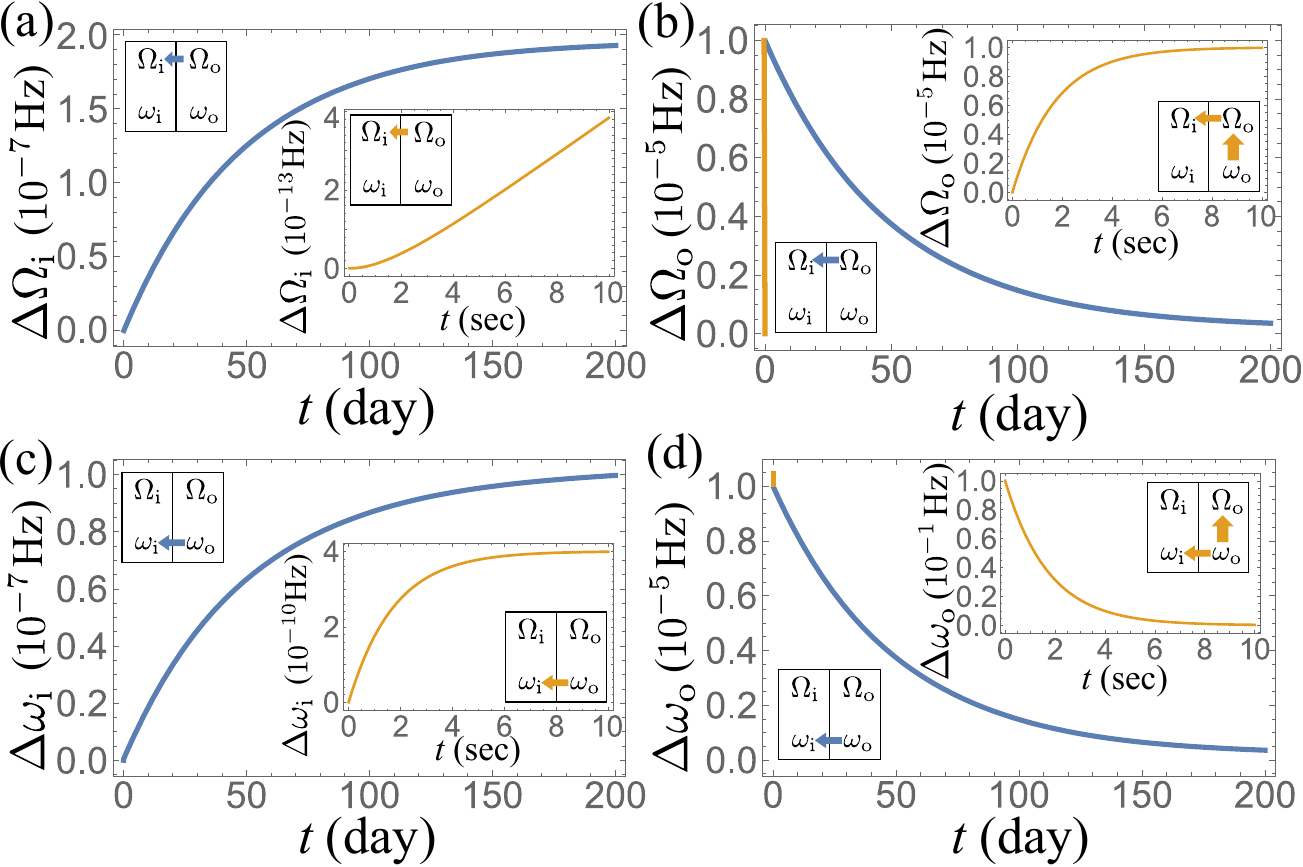}
 \caption{
Gyromagnetic glitch in the four-angular-velocity model.
The panels show the angular velocities as in Fig.~\ref{s-IP_G}, but with a larger $\alpha_{\rm o}=10^{-4}$ and initial conditions $\Delta\omega_{\rm o} = 0.1{\rm Hz}$ and $\Delta\Omega_{\rm o}=\Delta\Omega_{\rm i}=\Delta\omega_{\rm i} = 0{\rm Hz}$.
The crust microrotation, initialized with $\Delta\omega_{\rm o} = 0.1{\rm Hz}$, induces a rapid increase in $\Delta \Omega_{\rm o}$ 
[orange line in (b)], which we interpret as a glitch triggered by the gyromagnetic effect.
The rise time in our results is on the order of seconds, which agrees with observations~\cite{Ashton}.
}
 \label{b-IP_G-o}
\end{figure}
%%%%%%%%%%%%%%%%%%%%%%%%%%%%%%%%%%%%%%%%%%%%%%%%%%%%%%%
%
In Fig.~\ref{b-IP_G-o},
we demonstrate the case (ii), where
the glitch is directly induced by gyromagnetic angular momentum interconversion.
The unpinning and re-pinning mechanism itself is thought to be a non-equilibrium process, making it nontrivial to understand how angular momentum is transported through it.
Therefore, in the present case, we consider that the microrotations in the crust accumulate the angular momentum at first driven by the unpinning and re-pinning mechanism: the initial condition set as  
$\omega_{\rm o} > \omega_{\rm i} = \Omega_{\rm o} = \Omega_{\rm i}$. 
The moment of inertia of the fluid element is set to a large value $\alpha_{\rm o}=10^{-4}$ to focus on the situation where microrotation provides feedback to the angular velocity of the fluid~\cite{supplemental,Garcia,Rosswog}.
In this case, the angular momentum from the core is transported through the microrotation flows; thus, we refer to this process as the {\it Einstein–de Haas effect} or simply as the {\it gyromagnetic glitch}, shown in Fig.~\ref{b-IP_G-o}.
We find that the crust angular velocity undergoes a rapid increase over a very short time, a few seconds, followed by a rapid decrease of crust microrotation as shown in Fig.~\ref{b-IP_G-o}.
Notably, this time scale aligns with the observations reported in~\cite{Ashton}.
This rapid change suggests that a sudden shift in the crust microrotation could manifest as a glitch in the observed crust angular velocity, offering an alternative scenario of glitch caused by the gyromagnetic effect.

{\it Discussion.---}
The idea that the glitches of NSs may be related to the superfluidity of the internal matter has long been suggested, based on inferences from the relaxation time. Nevertheless, this Letter is the first to point out the connection between spin currents, which are closely related to superfluidity, and angular momentum transport, and to provide a qualitative formulation.
Overall, whether the moment of inertia of the fluid elements is large or small, there is a significant possibility that the outer spin could be strongly polarized via the Barnett effect.
In cases where the moment of inertia is large, the spin may significantly influence the flow dynamics, and observed changes in the outer angular velocity could, in fact, be reflecting underlying changes in spin rather than purely macroscopic effects.

With the simple model presented in this Letter, it is beyond our current capabilities to make detailed predictions about observable phenomena, including radiation from the NS magnetospheres
resulting from the microrotation. However, if extreme phenomena such as spin alignment within the crust occur via the Barnett effect or Einstein-de-Haas effects, they could have an impact on observables such as the photon polarization, 
and surface luminosity.
The large spin polarization can also lead to the triplet neutron superfluid in the crust~\cite{PhysRevC.108.L052802}.
While it is challenging to detect them statistically at present since the duration of glitches is quite short, we look forward to advances in observational techniques in the future.
In this Letter, we provide just two parameter sets that are consistent with the observed glitch phenomena. It should be noted that, with other parameter sets, it is also possible to explain anti-glitches~\cite{supplemental}.

{\it Conclusion.---}
In this Letter, we proposed a novel model for gyromagnetic angular momentum interconversion in NSs, focusing on macroscopic and microscopic angular momentum interactions in the crust and core. Using an extended hydrodynamics framework, we developed a four-angular-velocity model, going beyond the conventional two-angular-velocity approach.
Our analysis showed that %macroscopic dynamics in the crust can significantly influence the microscopic angular momentum,
the interconversion between macroscopic and microscopic angular momenta leads to
nontrivial effects such as neutron spin polarization and the occurrence of glitch/anti-glitch-type angular momentum transfer through gyromagnetic processes. 

These findings enhance our understanding of non-equilibrium angular momentum dynamics in NSs and highlight the significance of gyromagnetic effects in astrophysical phenomena, encouraging further theoretical, experimental, and observational research.

\begin{acknowledgments}
M. M. is grateful to Kazuya Harii for the valuable discussions during the early stages of this work.  
We acknowledge JSPS KAKENHI for Grants 
(Nos.~JP21K03436, 
 JP22K13981, 
 JP23H01839, 
 JP22K03681,
 JP23K22429,
JP23K22538, 
JP23K20841, 
 JP24K07054,
 and JP24H00322). 
 Y.~S. is supported by Pioneering Program of RIKEN for Evolution of Matter in the Universe
(r-EMU), the RIKEN TRIP initiative (RIKEN Quantum), and by JST ERATO Grant Number JPMJER2302, Japan.
M. M. is supported by the National Natural Science Foundation of China (NSFC) under Grant No. 12374126 and
by the Priority Program of Chinese Academy of Sciences under Grant No. XDB28000000.
\end{acknowledgments}

\bibliography{masterbib}%.bib}

\end{document}